# Differentiation of subjects of the Russian Federation according to the main parameters of socio-economic development

Natalia A. Sadovnikova[1], Leysan A. Davletshina[2], Olga A. Zolotareva[3], Olga O. Lebedinskaya[4]*

*This research was performed in the framework of the state task in the field of scientific activity of the Ministry of Science and Higher Education of the Russian Federation, project "Development of the methodology and a software platform for the construction of digital twins, intellectual analysis and forecast of complex economic systems", grant no. FSSW-2020-0008.*

**Abstract:** This article presents the results of a cluster analysis of the regions of the Russian Federation in terms of the main parameters of socio-economic development according to the data presented in the official data sources of the Federal State Statistics Service (Rosstat). Studied and analyzed the domestic and foreign (Eurostat) methodology for assessing the socio-economic development of territories. The aim of the study is to determine the main parameters of territorial differentiation and to identify key indicators that affect the socio-economic development of Russian regions. The authors have carried out a classification of the constituent entities of the Russian Federation not in terms of territorial location and geographical features, but in terms of the specifics and key parameters of the socio-economic situation.

## Introduction

The need to classify regions according to various factors is determined by the fact that the Russian Federation has a huge territory, in some parts of which there are distinctive climatic, demographic and socio-economic features, which is manifested in the differentiation of regions according to various indicators. The study and analysis of individual parameters of the socio-economic development of territories will not allow getting an understanding of the features of the development of subjects, only the use of the main indicators makes it possible to conduct a comprehensive assessment of differentiation in the territorial context.

The state regional policy is aimed at ensuring a balanced socio-economic development of the constituent entities of Russia. One of the most important goals of the Concept of long-term socio-economic development of the Russian Federation for the period up to 2020 is to reduce the level of interregional differentiation in the socio-economic state of regions and the quality of life. According to the presidential decrees in May, one of the goals of the country's sustainable development until 2024 is to increase the real incomes of citizens of all constituent entities of Russia, which largely determines the standard of living of the population. However, one can recall the opinion of N.V. Zubarevich that achieving equality of all regions is an almost impossible task due to too strong differences in the resources and potential of the regions, their climatic and territorial characteristics, population, etc. Therefore, it is important to emphasize that the task is to reduce regional differentiation, and not to completely eliminate it.

[1] Candidate of Economic Sciences, Professor, Head of the Department of Statistics, Russian Economic University named after G.V. Plekhanov, Sadovnikova.NA@rea.ru
[2] Candidate of Economic Sciences, Associate Professor of the Department of Statistics, Russian Economic University named after G.V. Plekhanov, Davletshina.LA@rea.ru
[3] Candidate of Economic Sciences, Associate Professor, Associate Professor of the Department of Statistics, Russian Economic University named after G.V. Plekhanov, Zolotareva.OA@rea.ru
[4] Candidate of Economic Sciences, Associate Professor, Associate Professor of the Department of Statistics, Russian Economic University named after G.V. Plekhanov, Lebedinskaya.OG@rea.ru

The uneven development and specificity of the sectoral structure of the economy of individual regions of Russia determine the presence of differences in the socio-economic development of territories. Achievement of target indicators stipulated by state programs presupposes the development and implementation of comprehensive measures in regions that are similar in various socio-economic parameters. In this regard, it becomes necessary to form stable groups (clusters) of regions.

In scientific publications, there are various methods of grouping regions according to the level of socio-economic development. A widespread approach is based on the calculation of the integral index, in accordance with which the regions are grouped. This method is used both in international and domestic practice. An alternative to the index method is the multivariate statistical approach, which is based on methods of cluster analysis. According to the purpose of the study, cluster analysis seems to be the optimal segmentation method, since the method has a variety of tools that allow us to assess the quality and stability of the formed groups.

In international practice, the European Observatory on Clusters and Industrial Change (EOCIC) has a special place in working with clusters. It provides political support for existing or new cluster initiatives at the national and regional levels. This definition is achieved through conceptual schemes and descriptions of modern cluster policy, which contributes to regional structural changes and the development of new industries.

EOCIC helps European regions and countries develop more effective and evidence-based cluster policies and initiatives to:
• developing world-class clusters with competitive value chains that span all sectors;
• support of industrial modernization;
• development of entrepreneurship in emerging industries with growth potential;
• expanding access of SMEs to clusters and internationalization of activities, etc.

EOCIC builds on the work of the European Cluster Observatory and the European Service Innovation Center, with a strong focus on industrial change and its key drivers such as service innovation, entrepreneurship, key enabling technologies, digitalization, creativity and eco-innovation, and resource-saving solutions ... A set of indicators drives each of these dimensions with a specific focus on triggering and zooming in.

EOCIC is considering supporting clusters in 29 European countries, 49 European regions, and 10 countries outside Europe (Brazil, Canada, China, Israel, Japan, Mexico, Singapore, South Korea, Taiwan and the United States). An in-depth analysis of a wide range of cluster programs in Europe and beyond shows what goals are being pursued by cluster programs in different parts of the world and how they are being achieved. In general, the analysis of EOCIC activities shows that cluster policy is widespread and that cluster support is an essential tool for economic development in Europe and around the world.

EOCIC uses a data visualization tool called Cluster Mapping, which shows the result of cluster mapping in a regional ecosystem scoreboard. It displays industry, cross-industry and regional indicators of industry clustering and their size, business performance and other statistical information.

Thus, clustering as a tool for assessing regional differentiation is relevant both in domestic and foreign practice. This study will use the main parameters of the socio-economic development of territories and their distribution into characteristic, similar groups (clusters).

**Methodology**

At the initial stage, it is necessary to determine a list of indicators on the basis of which the grouping of regions will be carried out. The feature space of cluster analysis is formed on the basis of the system of indicators of socio-economic development of territories used in the work (Fig. 1).

The system of statistical indicators proposed by the team of authors characterizing the main parameters of the socio-economic development of the territory is compiled on the basis of the following principles: simplicity, brevity and compactness. Thus, in the presented in Fig. 1 system of indicators highlighted the enlarged groups that characterize the relevant areas of socio-economic development of the territory: economic (macroeconomic indicators), demographic (characterizing the population and its change), standard of living and labor market.

**Results**

   *1. Composition of clusters by main indicators of socio-economic development of Russian regions*

The classification of the regions of the Russian Federation into qualitatively formed homogeneous groups makes it possible to identify regional differentiation, as well as to identify characteristic groups based on actual data on the main indicators of the socio-economic development of regions. In the process of conducting cluster analysis, much attention was paid to such issues as: complex classification of regions, identification of similarities and differences between them, inconsistency of classifications of the same objects on several grounds, the nature of the solution of these contradictions.

In the study, for the purpose of conducting cluster analysis, the official statistical data presented in the open resources of the Federal State Statistics Service (Rosstat) for 2018 were used (for a number of indicators, data for 2019 are not available at the time of the study). All 85 constituent entities of the Russian Federation are included in the clustering procedure.

Cluster analysis of the country's regions in terms of socio-economic development was carried out using the SPSS Statistics and R.

Before starting the analysis itself, it is necessary to determine the number of clusters, i.e. the number of groups into which the regions of Russia will be divided. In order to determine the number of clusters for clustering by the k-means method, it was decided to carry out calculations in the R program using the elbow method. This method considers the nature of the change in the spread of Wtotal with an increase in the number of groups k. Combining all 85 observations into one group, we have the largest intra-cluster variance, which will decrease to 0 as k → nk → n. At a certain stage, one can see that the decrease in this variance slows down - on the graph, this occurs at the point called the "elbow" (Fig. 2).

Further, knowing the number of clusters, a cluster analysis was carried out by the k-means method using the SPSS Statistics application package. We choose Euclidean distance as a metric. Let's construct a grouping in a complete feature space. Clustering of the constituent entities of Russia by the nearest neighbor method also allowed us to conclude that the studied administrative-territorial units should be organized into four clusters (Fig. 3).

The dimensions of the clusters presented in Table 1 are different. Thus, the largest number of regions is included in cluster 2 (39 subjects of the federation). In this cluster, there are subjects from 7 federal districts, only representatives of the North Caucasian Federal District are absent. The next largest cluster - 1, includes 33 regions of Russia, representing all 8 federal districts of the country. Clusters 3 and 4 differ significantly: they include 4 and 9 subjects of the federation, respectively. Cluster 3 includes regions that are representatives of the Central and Ural Federal

Districts, in fact, including the city of Moscow and the Tyumen Region with two autonomous districts. A distinctive feature of cluster 4 is that it includes national republics from the South, North Caucasian and Siberian federal districts.

Thus, 85 constituent entities of the Russian Federation, according to the main indicators of socio-economic development, according to official statistics for 2018, according to the system of indicators used in this study, after applying the clustering procedure, were divided into 4 groups with similar content. Structurally, the largest, the second cluster includes 45.8% of the country's regions, the first cluster - 38.7% and the smallest clusters are three and four, 5% and 10.5%, respectively.

   *2. Average values of factors for enlarged groups and their analysis*

In order to better understand and analyze the above clusters, average values are calculated.

The enlarged groups of indicators include the main indicators characterizing the socio-economic development of the territory: economic (GRP per capita, rubles; The volume of investments in fixed assets, million rubles; The cost of fixed assets, million rubles; Costs of technological innovation, million rubles; Industrial producer price index,%), demographic (Population change,%; Natural population growth rates per 1000 people; Migration growth rates per 10,000 people; Demographic load factors, per 1000 people of working age; Expected life expectancy at birth, years), standard of living (Average per capita cash income, rubles; The share of the population with cash incomes below the subsistence minimum, in%; Gini coefficient, in times; Consumer price index,%; The cost of a fixed set of consumer goods and services, rubles) and the labor market (number of labor resources, thousand people .; Share of people under working age employed in the economy in the total number of employed,%; Unemployment rate, %; Real accrued wages of employees of organizations,%).

Characterization of the main parameters of the economic development of the subjects of the Federation makes it possible to assess the activity and involvement of territories in the economic life of the country, their comparison and comparison (Table 2).

Analyzing the average values of the main indicators characterizing the economic development of territories by clusters, the following feature is highlighted: all the maximum values of the enlarged group under consideration are characteristic of cluster 3, which includes the city of Moscow, Tyumen region (without AO), Khanty-Mansiysk Autonomous Okrug, Yamalo-Nenets Autonomous Okrug, and all the minimum values are inherent in cluster 4 (Republic of Kalmykia, Republic of Dagestan, Republic of Ingushetia, Kabardino-Balkar Republic, Karachay-Cherkess Republic, Republic of North Ossetia-Alania, Chechen Republic, Altai Republic, Tyva Republic). The differences between the clusters are significant. So, for example, GRP per capita according to the average values of the third cluster is 56 times higher than the average values of cluster 4, and the cost of technological innovation is 687 times!

The significant differences in the average values for the main indicators characterizing economic development for clusters 3 and 4 can be explained primarily by the composition of regions in clusters.

Cluster 3 includes regions with abnormally high economic indicators: Moscow, the capital of Russia, a city of federal significance, the largest financial center on a national scale and a center for managing a significant part of the country's economy. For example, more than half of the banks registered in the country are concentrated in Moscow, while they account for 90% of banking assets. In addition, most of the largest companies are registered and have their central offices in Moscow, although their production can be completely located thousands of kilometers from the capital. As of November 2019, 104 of the country's 200 largest enterprises are registered in Moscow.

In the Tyumen region (including the AO) in recent years, the economic growth rate has become one of the highest among the regions of Russia. The region ranks first in the country in terms of industrial output. The main industry of specialization is oil and gas production, concentrated mainly in the Khanty-Mansi Autonomous Okrug-Yugra and Yamalo-Nenets Autonomous Okrug. In the south of the Tyumen region there are petrochemical enterprises, mechanical engineering (production of oil field, geological exploration, oil refining equipment), agro-industrial complex (production of meat and dairy products and vegetables), scientific and educational institutions.

Cluster 4 includes most of the subjects of the North Caucasus Federal District (with the exception of the Stavropol Territory), the regions of which are subsidized. So, in 2019 in Dagestan, subsidies accounted for 52% of the revenues of the consolidated budget of the region, in Chechnya - 50%, in Ingushetia - 49%, in Kabardino-Balkaria - 35%, in Karachay-Cherkessia - 34%. The Republic of Tyva became the most dependent on federal subsidies - subsidies accounted for 54% of the region's consolidated budget revenues. Historically, the regions represented in cluster 4 have not had large-scale industries or mining operations. The most traditional for these republics are agriculture (livestock and crop production), trade and government. With the exception of the Republic of Tuva. In this region, the first place in terms of the main types of economic activity is mining.

Closer to the maximum values are the average values of cluster 1 (33 subjects of the federation). The cluster includes the most successful regions of each of the federal districts of Russia - all the average values of key indicators characterizing the economic development of territories are located immediately after the leader of the group, ranking second in all positions. In 2018, the average value for the cluster of the gross regional product per capita is 92,101.3 rubles, which is 6 times less than the average value for cluster 3 (the highest values for this group of indicators), but at the same time, the presented value is more than 9 times more than the average for cluster 4 (the lowest values for this group of indicators). The average value for the cluster of the cost of fixed assets is 1,744,826.3 million rubles, which is almost 8 times less than the average value for cluster 3 and 4.5 times more than the average value for cluster 4. It should be noted that the smallest differences are characteristic of the costs of technological innovation - the difference with the highest average values of the 3rd cluster is 3.2 times. According to this indicator, there are tangible differences in the average values for clusters: the highest values are in cluster 3, cluster 1 is next, then cluster 2, whose values are 25 times less than cluster 3 and almost 8 times less than cluster 1, and 4 is an indisputable outsider in the analyzed group. cluster - 687 times less than the average values of cluster 3, 215 times less than cluster 1 and 27 times less than cluster 2. Thus, the regions included in cluster 1 are characterized by significant economic activity, significant volumes of investments in fixed assets, developing technological innovations.

The largest cluster 2 (39 subjects) included those regions that have not yet entered the group of economic outsiders, but at the same time do not have particularly high values in terms of the average values of key indicators characterizing the economic development of territories - for five analyzed In terms of parameters, cluster always ranks third behind the recognized leader (cluster 3) and cluster 1, but ahead of the least economically prosperous cluster 4. For example, the average cost of technological innovation in cluster 2 is 27 times higher than the similar values of cluster 4. At the same time, there are practically no differences in the average values of the cost of fixed assets (about 15%); in the regions included in clusters 2 and 4, the cost of fixed assets is extremely low, which may indicate both a significant deterioration of fixed assets and their obsolescence.

Summarizing the results of the cluster analysis for an enlarged group of indicators characterizing the main parameters of the economic development of regions, the following conclusions can be drawn. First, the system of statistical indicators developed by the authors and used in the work carries the task of analyzing the key parameters of socio-economic development and does not include an extended list of indicators. Secondly, the results of the clustering procedure carried out made it possible to group the regions of the country according to similar characteristics. Thirdly, from the 4 obtained clusters, a small leader cluster (cluster 3) and an outsider cluster (cluster 4), including 4 and 9 subjects of the federation, respectively, stand out. Fourth, the most voluminous clusters 1 (33 subjects) and 2 (39 subjects) characterize about 85% of the country's regions, each of the clusters has its own peculiarity. Fifthly, for a more detailed disclosure of the processes occurring at the regional level, the directions and speed of changes, it is necessary to continue working on the topic, including the study of both time series and expanding the system of indicators.

The next enlarged group of indicators includes key parameters that characterize the standard of living of the population. In modern conditions of economic development, the concept of "living standards of the population" is of particular importance, since it most reflects the development of health care, education and other spheres of the country's life. The standard of living is a complex and multifaceted category that requires statistical analysis taking into account the multifactorial and dynamic nature of socio-economic processes.

According to the Convention of the International Labor Organization (hereinafter referred to as the ILO) "On the main goals and norms of social policy", an individual has the right to the standard of living that is necessary to maintain the health and well-being of himself and his family at a decent level, as well as the right to safety in case of loss of livelihood in circumstances beyond his control. As the United Nations (hereinafter referred to as the UN) recommends, a person's standard of living should be assessed by a system of indicators that indicate health, consumption, employment, education, housing, social security and other characteristics.

The standard of living of the population is a complex socio-economic category that characterizes the level of satisfaction of the needs of the population in tangible goods and intangible services, as well as the conditions in society for the development and satisfaction of these needs. In the study of the socio-economic development of Russian regions, the clustering procedure includes five key indicators (Table 3).

The average per capita monetary income of the population in 2018 in Russia as a whole amounted to 33,178 rubles. This figure is 23,737.8 rubles. (71.55%) less than the average income of the population of the four regions of the richest cluster 3 and higher than the same indicators for subjects from the other three clusters. The most approximate are the values of average per capita money income for cluster 1 - the average values for the cluster are lower than the national ones by 1,093.2 rubles. or 3.3%. The most tense situation was recorded in the regions of cluster 4 - the average values for the cluster are lower than the national ones by 13,244 rubles. or 40%, which is quite expected for the values that were identified in the enlarged group of indicators characterizing the economic development of territories. The average per capita money income for cluster 2 is lower than the national average, and the average for clusters 1 and 3, but higher than cluster 4.

In 2018, in the Russian Federation, about 12.60% of the population had cash incomes below the subsistence level, which is 3.5 pp. (by 27.8%) higher than the poverty level in the richest regions of Russia (cluster 3). The volumes of the population with monetary incomes below the subsistence level in cluster 1 are closest to the federal ones - 12.0%, which is 4.8% lower. The average value of the indicator for cluster 2 and 4 is higher than the national values by 16.7% and 84.1%, respectively. Thus, the most tense situation with the level of poverty in the regions included in cluster 4 - almost a quarter of the population lives below the poverty line.

Economists believe that the Gini coefficient should not be higher than 0.3-0.4. When the index is higher, there is high inequality in the country. It slows down the pace of economic development and forms a "poverty trap" in which society becomes poorer with each generation. The degree of differentiation of the Russian population in terms of money income was high. The Gini coefficient in 2018 was 0.413 times, which exceeds the same indicators for the regions of all four clusters. The highest average value of the coefficient for the regions of Russia in 2018 corresponds to cluster 4, the most economically prosperous - 0.410 times. Thus, high economic indicators and the highest average per capita money incomes do not guarantee income equality in society. The rest of the clusters, despite significant differences in the previous socio-economic indicators, have normal values of the Gini coefficient.

The consumer price index in the analyzed period for the country as a whole was 104.3%, an absolutely identical average value is characteristic of cluster 1. In the regions of cluster 2, prices in 2018 grew more intensively than the national ones by 0.4 pp, in regions of cluster 3 and 4 the average rate of price growth is lower than in Russia as a whole.

The cost of a fixed set of consumer goods and services in the Russian Federation in 2018, according to Rosstat, was 15467.9 rubles, which is 3558.1 rubles (or 23%) below the average value of the most economically prosperous regions of cluster 3. The regions have a slight advantage. included in cluster 1 (60.8 rubles). The lowest average values of the cost of a fixed set of consumer goods and services in the regions included in cluster 4.

After analyzing the average values of key indicators characterizing the standard of living of the population in 2018 for each cluster and comparing them with the all-Russian values, we can formulate the following conclusions. First, in the considered enlarged group of indicators, the leader remained unchanged - the regions included in cluster 3 are invariably in the first positions (with the exception of income differentiation). The outsider has not changed either - the republics of cluster 4, with the exception of income differentiation (in cluster 4, the lowest value of the coefficient) and the consumer price index (below the all-Russian values) by key parameters characterizing the standard of living of the population (average per capita money income, poverty level, cost of a fixed goods and services) significantly lag behind both the average values of other clusters and the national scale. Third, the average values of the regions included in cluster 1 are the closest to the all-Russian values - deviations of all the indicators under consideration are within 5%.

Labor market statistics are a very important tool for the development of economic and social policies of states, as well as for analyzing the structure of the population. Labor market statistics consider the size of the employed and unemployed population as two components of the labor force, based on the measurement of the labor force, monitoring is carried out and strategies are developed to increase the employment of the population providing their labor for the production of goods and the provision of services.

The size and composition of labor resources, labor force, employed and unemployed population is dynamic and changeable. Understanding the prevailing trends in the indicators under consideration is possible only when reliable, detailed and versatile information is used.

Four key indicators were used as the main indicators characterizing regional labor markets in order to characterize territorial differentiation.

The number of labor resources is an indicator characterizing the "free" labor reserves not participating in the regional economy, which, if necessary, can be attracted. The highest average values for clusters are typical for the third cluster - 3633.4 thousand people, the smallest - for the fourth cluster, 7.3 times lower than the values of the first. When comparing with the leader, it was revealed that the average values of the indicator in the regions included in the first cluster are closer to the leader, but nevertheless lag behind in values 2.4 times, the second cluster - 6 times. Thus, the regions included in the third cluster have the largest labor reserve, the most tense situation with free labor resources in the republics included in the fourth cluster. To understand such a distribution, in addition to indicators characterizing the labor market, it is necessary to study the changes and structure of the population in the regions.

The share of people under working age employed in the economy in the total number of employed - shows how much the number of labor resources in the regions is formed at the expense of young people who, for social and economic reasons, are forced to work outside of working age. The most tense situation in this case was formed in the regions included in the second cluster - there 12% of the employed were formed at the expense of persons younger than the working age. The most favorable situation is in the regions of the first cluster - about 1% of those employed in the economy are formed by people under the working age.

The unemployment rate is a relative value that characterizes the ratio of registered unemployed to the total labor force, and represents a socio-economic phenomenon in which a part of the population capable and ready to work cannot find a job. The most acute problem is registered unemployment in the republics included in the fourth cluster. Such a distribution is quite expected, given the values presented in the group of main indicators characterizing the economic development of territories.

The real accrued wages of employees of organizations reflects the relative change in nominal wages, taking into account changes in prices in the current period. The maximum average values of this indicator are inherent in the fourth cluster - in this case, we are talking more about a high share of government support measures, including social transfers, rather than about significant economic growth or an increase in wages of regional workers.

The labor market is one of the elements of a market economy. It is a system of public relations in the coordination of the interests of employers and hired labor. According to the key parameters included in the system of indicators that characterize the regional labor market and for the four clusters formed in the regions of Russia in 2018, the following is observed: a significant level of unemployment with the least number of labor resources, with the maximum value of real accrued wages, is characteristic of the fourth cluster. The third cluster, being the leaders in the previous enlarged groups of indicators, in the analysis of the labor market also occupies the leading places - the smallest scale of the unemployment rate and the attraction of people younger than the working age to work and the greatest value of labor resources with a minimum increase in real gross wages. As before, the average values for the first cluster are closer to the leader (cluster 3), and the average values of the second cluster are closer to the outsider.

The demographic situation is a complex quantitative characteristic and qualitative assessment of the demographic processes occurring in a certain territory: their trends, outcomes for a certain period and consequences. It is formed due to many different indicators and that is why it is rational to analyze it on the basis of a certain list of sequential and logically interrelated tasks, the solution of which allows you to get a complete and comprehensive idea of the actual state and development prospects of the research object. In this study, the main parameters that characterize the demographic situation will be used (Table 5.).

In the economic and statistical study of the socio-economic situation, both at the federal and regional levels, an important role belongs to the analysis of population dynamics, which traditionally represents the overall results of the development of society. This step allows us to judge the nature of the transformation of administrative-territorial entities.

The change in the population size in 2018 for the analyzed 4 clusters is as follows: in the regions included in clusters three and four, the population increases by 0.7% and 0.4%, respectively, while in the largest first and second clusters (84.5% of the subjects of the Federation in total) recorded an average decrease in the population. Such transformations occur due to the predominance of negative phenomena, such as the preponderance of mortality over births or the greater importance of the number of those who left over those who arrived. The analysis of indicators in absolute values for the purpose of their further comparison is incorrect, since the population size in the regions is different. In the work, in order to determine the factors of change in the population size, we used relative values - the coefficients of natural and migratory population growth.

From the point of view of structural changes, the analyzed clusters have a very curious form. The average values of the main demographic indicators in the regions included in the first cluster were distributed as follows: natural population decline (3 per 1000 people of the population) versus migration growth (17.1 per 10,000 people) ultimately gives a decrease in the population, i.e. ... the intensity of natural loss is higher and is not covered by migration gain.

The average values of the main demographic indicators in the regions included in the second cluster are unidirectional, have negative values - both natural and migration losses in total give the largest population decline among the analyzed clusters (0.79%), i.e. on average, these regions are characterized by high mortality rates with migratory unattractive territories.

The average values of the main demographic indicators in the regions included in the third cluster are also unidirectional - both natural and migration growth, as a result of the highest values of population growth in the cluster environment (0.70%).

A feature of the average values of the main demographic indicators in the regions included in the fourth cluster is the fact that the birth rate in these subjects is such that it not only compensates for mortality, but also for migration decline, thereby ensuring an increase in the population of the territories, i.e. these regions act as donors for the rest.

Demographic load ratios make it possible to determine the existing proportions in the age structure of the population: the ratio of disabled people per 1000 people of working age. The largest value of the coefficient falls on the average value of the second cluster (835.9 per 1000 people of working age), the lowest value is typical for the third cluster - 669 people of disabled age per 1000 people of working age.

Life expectancy at birth is the number of years that, on average, would have to live for a person from the generation of births, provided that throughout the life of this generation, the age-specific mortality remains at the level of the year for which the indicator is calculated. Life expectancy is the most adequate generalizing characteristic of mortality. The highest values are in the republics included in the fourth cluster (75.1 years), and the lowest in the regions of the second cluster - 71.4 years.

Thus, after analyzing the main parameters of the demographic development of Russian regions, the following features of the territories were identified. First, from the point of view of changes in the population size and its constituent elements, all clusters are different (due to the directions of changes). Secondly, in the second cluster, which has negative values for the change in population, the expected high values of the dependency ratio. Thirdly, the highest values of life expectancy at birth are typical for the republics included in cluster 4, where 7 out of 9 are regions of the North Caucasus with national culture and traditions of longevity.

## **Conclusion**

This study was carried out within the framework of the state task in the field of scientific activity of the Ministry of Science and Higher Education of the Russian Federation on the topic

The authors had the goal of assessing the regional differentiation of the constituent entities of the Russian Federation by the level of socio-economic development, having developed an original system of statistical indicators that include the main parameters of the socio-economic development of territories and using the method of cluster analysis to identify typical groups of regions from the standpoint of the specifics and key parameters of the socio-economic situation ...

In the work, before compiling a system of statistical indicators, an analysis of scientific literature was carried out and the international practice of working with clusters was considered. The European Observatory on Clusters and Industrial Change provides policy support for existing or new cluster initiatives at national and regional levels. This was made possible by pilot regions in transition in industry, the cluster stress testing tool, and conceptual sketches and descriptions of modern cluster policies that facilitate regional structural change and the development of new industries.

The system of statistical indicators developed and proposed by the authors includes key parameters of the socio-economic development of territories and at this stage of the study does not pretend to be an all-embracing one, since the team of authors is faced with the task of assessing the key parameters of the socio-economic development of the subjects of the federation with the subsequent improvement and expansion of the assessment system.

Cluster analysis of Russian regions by the parameters of socio-economic development was carried out using the SPSS Statistics and R software packages. 85 constituent entities of the Russian Federation by the main indicators of socio-economic development after applying the clustering procedure were divided into 4 groups similar in content. Structurally, the largest, the second cluster includes 45.8% of the country's regions, the first cluster - 38.7% and the smallest clusters are three and four, 5% and 10.5%, respectively.

Summing up the results of regional differentiation in terms of the level of socio-economic development of the constituent entities of the Russian Federation in 2018, it should be noted that in the course of analysis and interpretation of the results, clusters with leading values (cluster 3) and significantly lagging (cluster 4), as well as cluster 1, were identified, whose values were always closer to cluster 3 and all-Russian.

This topic is relevant, has an applied nature and requires a more detailed disclosure of the processes occurring at the regional level, the directions and speed of changes. The team of authors considers it necessary to continue work on the topic by including the study of both time series and expanding the system of indicators.